\documentclass[acmsmall]{acmart}
\usepackage{enumitem}
\usepackage[normalem]{ulem}
\usepackage{caption}
\useunder{\uline}{\ul}{}

\AtBeginDocument{%
  \providecommand\BibTeX{{%
    \normalfont B\kern-0.5em{\scshape i\kern-0.25em b}\kern-0.8em\TeX}}}

\newcommand{\parabold}[1]{\vspace{0.5ex}\noindent\textbf{#1.}}

\definecolor{changes}{RGB}{0,0,0}
\begin{document}

\title[One Style Does Not Regulate All]{One Style Does Not Regulate All: Moderation Practices in Public and Private WhatsApp Groups}

\author{Farhana Shahid}
 \affiliation{
   \institution{Cornell University}
   \city{Ithaca}
   \country{United States}
   }
\email{fs468@cornell.edu}
\orcid{0000-0003-3004-7099}

\author{Dhruv Agarwal}
 \affiliation{
   \institution{Cornell University}
   \city{Ithaca}
   \country{United States}
   }
\email{da399@cornell.edu}
\orcid{0000-0002-1090-3583}

\author{Aditya Vashistha}
 \affiliation{
   \institution{Cornell University}
   \city{Ithaca}
   \country{United States}
   }
\email{adityav@cornell.edu}
\orcid{0000-0001-5693-3326}

\begin{abstract}
WhatsApp is the largest social media platform in the Global South and is a virulent force in global misinformation and political propaganda. Due to end-to-end encryption WhatsApp can barely review any content and mostly rely on volunteer moderation by group admins. Yet, little is known about how WhatsApp group admins manage their groups, what factors and values influence moderation decisions, and what challenges they face while managing their groups. To fill this gap, we interviewed admins of 32 diverse groups and reviewed content from 30 public groups in India and Bangladesh. We observed notable differences in the formation, members' behavior, and moderation of public versus private groups, as well as in how WhatsApp admins operate compared to those on other platforms. We used Baumrind's typology of \textit{`parenting styles'} as a lens to examine how admins enact care and control during volunteer moderation. We identified four styles based on how caring and controlling the admins are and discuss design recommendations to help them better manage problematic content in WhatsApp groups.

\end{abstract}

\ccsdesc[500]{Human-centered computing~Empirical studies in HCI}

\keywords{WhatsApp, content moderation, encryption, Global South}

\setcopyright{acmlicensed}
\acmJournal{PACMHCI}
\acmYear{2025} \acmVolume{9} \acmNumber{2} \acmArticle{CSCW144} \acmMonth{4}\acmDOI{10.1145/3711042}

\received{January 2024}
\received[revised]{July 2024}
\received[accepted]{October 2024}

\maketitle
\section{Introduction}


WhatsApp is the most popular instant messaging platform with the second largest active social media userbase, majority of whom are from the Global South~\cite{Morley-2022, Lua-2023}. WhatsApp promotes its \textit{`groups'} as an effortless way to bring together one's closest people. Indeed, most private WhatsApp groups consist of family, friends, or neighbors, while public groups involve strangers~\cite{Melo-2019, Gillespie-2020}. Although all groups start \textit{private}, when the group creator or manager, usually known as \textit{admin}, shares the invite link online for anyone to join, the group is considered \textit{public}. 

Due to end-to-end encryption WhatsApp has limited oversight on user-generated content. Thus, misinformation, hate speech, and targeted polarization run rampant on WhatsApp~\cite{farooq-2017, Machado-2019, Garimella-2020, Evangelista-2019, Jakesch-2021}, fueling mob lynching and communal violence in many Global South countries~\cite{banaji-2019, Khandelwal-2022}. WhatsApp uses unencrypted user metadata, low-paid human moderators, and AI algorithms to detect abusive behavior and disable suspected accounts~\cite{child-abuse}. However, these measures are often sporadic and insufficient to prevent the spread of harmful content~\cite{Constine-2018}. Due to encryption, paid moderators can only review messages that are reported by users~\cite{Elkind-2021}. Privacy concerns~\cite{wang-2023} and the reluctance to report misinformation in close-knit groups~\cite{Varanasi-2022, Kevin-2022} practically limit the scope of content that paid moderators can review. Therefore, WhatsApp relies on the volunteer labor of admins to keep the groups safe for all members~\cite{community-mod}. \color{changes}On WhatsApp, whoever creates a group is assigned the role of ``admin'' by default. Only existing admin can make other group members admin if needed. Although group members might call out when people share problematic content in the group, WhatsApp has rolled out moderation features (e.g., deleting messages or removing users from the group) only for group admins~\cite{admindelete-2022}.\color{black}

 A large body of HCI and CSCW scholarship has examined the roles of volunteer moderators in Facebook groups, subreddits, and Twitch channels through the lenses of community ownership, care, and control~\cite{Wohn-2019, Seering-2019, Juneja-2020, Sultana-2022, Seering-2022}. However, WhatsApp is substantially different from other platforms because it is end-to-end encrypted and often connects people, who already know each other \textit{offline}. The existing offline ties with group members require care on admins' part that go beyond managing the group, which usually involve some form of control. Yet, little is known about the roles of admins in WhatsApp groups and what challenges they face while managing their groups. To fill this critical gap, we examine admins' roles and moderation practices in two demographics in the Global South: India, which has the world's largest WhatsApp userbase~\cite{wastat-2023} and Bangladesh. 
In particular, we ask the following questions:

\begin{itemize}
    \item[\textbf{RQ1:}] How do admins in India and Bangladesh exercise care and control in response to problematic content shared in public and private WhatsApp groups?
    
    \item[\textbf{RQ2:}] How can WhatsApp improve volunteer content moderation in public and private groups? 
\end{itemize}

To find answers, we observed user activities in 30 public WhatsApp groups and interviewed 32 admins from different public and private groups in India and Bangladesh. To examine how care and control play out during volunteer moderation on WhatsApp (RQ1), we build upon psychologist Diana Baumrind's work on the dynamics between care and control in case of parenting~\cite{baumrind1991}. Baumrind identified four parenting styles based on the degree of parental care and control, which are: authoritative, authoritarian, permissive, and uninvolved. Considering the parallels between the responsibilities of parents in safeguarding children and volunteer admins in protecting community members, we apply Baumrind's typology to describe the various governance styles that we observed in private and public WhatsApp groups in India and Bangladesh. 

As observed in our sample, \textit{authoritative} admins upheld group's well-being, defined group rules, debunked misinformation, removed problematic content, and often banned the offenders. Unlike moderators on other platforms, authoritative admins were transparent about moderation due to the offline social ties they shared with group members. We observed authoritative approach mostly in private groups with weaker social ties (e.g., educational, professional, and organizational groups). Whereas, admins in private groups with strong social ties were \textit{permissive}, i.e., they ignored problematic content shared by family and friends to avoid jeopardizing their personal relations. In collectivist cultures like India and Bangladesh, offline relations shape who can or cannot be moderated in close-knit WhatsApp groups, which is unlikely on other platforms. In contrast, lack of social ties in public groups allowed admins either to fully neglect their groups or enforce harsh measures. For example, \textit{authoritarian} admins disabled messaging in public groups to prevent problematic content without caring for the communication needs of group members. Whereas, \textit{uninvolved} admins abandoned public groups and did not bother about the harmful content that circulated in their groups. 

We draw from our findings to answer RQ2 and stress the need to go beyond \textit{``one size fits all''} moderation tools, considering the strengths and weaknesses of different moderation styles grounded within Baumrind's typology. We recommend designs to empower admins while maintaining accountability and striking a balance in their power. This involves providing suitable avenues for authoritative and permissive admins to confront offenders, resources to educate group members against impulsive sharing of harmful content, and introducing measures that push authoritarian and uninvolved admins to responsibly engage with their groups. Overall, our work makes the following contributions to HCI and CSCW scholarship:

\begin{itemize}[leftmargin=1.75\parindent]
    \item A qualitative study built on Baumrind's typology, revealing the interplay between care and control during volunteer moderation in public and private WhatsApp groups compared to those on other platforms.
    \item A range of design recommendations to improve volunteer moderation on end-to-end encrypted platforms. 
\end{itemize}

\section{Related Work}

Social media platforms use different titles for users, who voluntarily manage online communities, e.g., \textit{admin} in WhatsApp and Telegram groups, \textit{moderator} in subreddits and Twitch channels, while Facebook groups have both. Irrespective of the title, admins and moderators play an important role in keeping the online community safe. 
We situate our work first by describing prior work in HCI and CSCW that extensively studies volunteer community moderation on Facebook, Reddit, and Twitch. We then discuss existing moderation practices on WhatsApp. Finally, we describe Baumrind's work on the measures of care and control in different `parenting styles'~\cite{baumrind1991} that we use as a lens to describe volunteer moderation practices in WhatsApp groups in India and Bangladesh. 

\subsection{Community Management and Volunteer Moderation} \label{subsec:moderators-other-platforms}
Admins and moderators on Facebook, Reddit, and Twitch often take pride in their position and feel a strong sense of community ownership~\cite{Malinen-2021}. Usually, moderators are existing users of the platform, who are either active community members, known to the existing moderators, or have prior experiences of moderation~\cite{Seering-2019}. In many Facebook groups, admins are elected by voting among interested users~\cite{Sultana-2022}. Many subreddits also have a formal application and interview process to appoint moderators~\cite{Matias-2019}. Admins and moderators mainly manage user-reported content, delete problematic content, mute or ban offenders, and resolve appeals~\cite{Seering-2019, Sultana-2022, Wohn-2019}. They use different auto-moderation tools and platform-provided features to prevent violations either by pre-approving posts or turning off post comments~\cite{Seering-2019, Sultana-2022}. Many Reddit and Twitch moderators use open-source bots to filter and flag posts that contain specific keywords~\cite{Seering-2019}. They often discuss moderation decisions with fellow admins and moderators and distribute the workload among themselves~\cite{cai-2021, Cai-2022, Sultana-2022}. However, they rarely engage with community members, answer their questions, or participate in conversations to avoid influencing group discussions with their positions of power~\cite{Wohn-2019, Malinen-2021}. 

Community moderators often lack transparency and accountability to users~\cite{Malinen-2021, Koshy-2023}. They enforce hidden rules and often remove content without any notification or explanation to prevent the gaming of community rules by the offenders and avoid disputes with them~\cite{Seering-2019, Wohn-2019, Juneja-2020}. Offenders usually respond to moderators' actions either by ignoring, improving their behavior, asking for clarification, or showing aggression~\cite{Seering-2019, Matias-2019}. Due to lack of training and onboarding processes, most volunteer moderators learn through `trial-and-error', i.e., they usually make rules or update existing ones following unexpected misconducts~\cite{Seering-2019, Wohn-2019}. On Reddit, moderators engage in reflective practices to update their mental models of community standards and moderation practices~\cite{Cullen-Kairam-2022}. Admins and moderators also rely on common sense, local laws, advice from fellow moderators, 
platform-provided resources, and help articles to learn about moderation~\cite{Seering-Kairam-2022, Sultana-2022}.  

Although these studies examine volunteer moderation on Facebook, Reddit, and Twitch, the findings may not apply to end-to-end encrypted platforms like WhatsApp, where user-generated content largely fall outside platform's oversight. Despite having large user base, volunteer moderation on WhatsApp is relatively understudied. Unlike other online communities, most WhatsApp group members share strong offline relation~\cite{Gillespie-2020} and there is limited understanding of how such dynamics might affect volunteer moderation. Our work addresses this gap by investigating moderation practices in various public and private WhatsApp groups in India and Bangladesh. 

\subsection{Content Moderation on WhatsApp}

Although WhatsApp is one of the main sources of harmful content propagating online~\cite{rossini2021, feng2022, Varanasi-2022}, end-to-end encryption makes it difficult for WhatsApp to review, moderate, and remove problematic content unless it is reported by users. Nevertheless, WhatsApp uses unencrypted account metadata (e.g., profile and group photos, account description) and in-app user reports to flag and ban abusive accounts~\cite{child-abuse, Gillespie-2020}. For example, WhatsApp banned more than 3 million user accounts in India in July 2023~\cite{WA-india-reports}. However, such measures often fail to prevent problematic content on WhatsApp. For example, users feel reluctant to report abusive messages containing sensitive personal information, fearing their personal data will be shared with external moderators~\cite{wang-2023}. Although there are some \textit{tiplines}, i.e., the WhatsApp accounts of fact-checking organizations that allow users to request fact-checks, these services 
remain underutilized because fact-checking is time-consuming, and most users find it exhausting to verify the bulk of messages they receive on WhatsApp~\cite{reis2020debunked, kazemi-2022, Kanthawala2022}. In the same vein, privacy-preserving measures, such as storing hashes of fact-checked misinformation on user devices, are ineffective in handling new abusive content that does not match with existing patterns~\cite{Reis-2020, kamara2022outside, reis2020debunked}. Moreover, limiting the number of forwards and labeling which messages are \textit{`forwarded many times'} barely stop the propagation of harmful content~\cite{deFreitasMelo2019}.

As these technological fixes fail, there has been an increased focus on user-driven social correction. Many studies examine how users' demographics impact their tendency to correct misinformation on WhatsApp and the results vary across different geographies. For example, male users tend to correct their peers more in WhatsApp groups in Brazil~\cite{Vijaykumar-2021} and Singapore~\cite{Kuru-2022}, whereas in Turkey, females are more likely to correct others~\cite{Kuru-2022}. In Brazil, older group members are more likely to confront misinformation~\cite{Vijaykumar-2021}, whereas in the US and Singapore, younger people are more likely to do so~\cite{Kuru-2022}. Moreover, highly educated users in Brazil participate in correcting misinformation more than others~\cite{Vijaykumar-2021}. Group dynamics and group ties also shape social correction in WhatsApp groups. A cross-cultural study in the US, Singapore, and Turkey revealed that users who have greater trust in group members are more likely to correct misinformation~\cite{Kuru-2022}. \citet{Pasquetto-2022} observed the effect of strong group ties in India and Pakistan, where users are more likely to share debunking messages that come from their close friends, family, or politically like-minded individuals. However, a study with WhatsApp users in the UK showed that irrespective of group ties, setting up informal group rules can foster collective reflection and epistemic vigilance against misinformation~\cite{Chadwick-2023}. 

Regardless of culture, community-driven corrections are considered stressful, particularly in tight-knit groups~\cite{Pasquetto2020mim}. \citet{scott-2023} studied family WhatsApp groups in the UK and found that family hierarchy, difficult personalities and biased opinions of misinformed family members, and lack of support to confront them online make it burdensome to challenge misinformation. Several studies with WhatsApp users based in India~\cite{Varanasi-2022, Malhotra-2023}, Singapore~\cite{Sheryl-2022}, and the US~\cite{feng2022} show that young group members usually ignore misinformation shared by elderly relatives to avoid conflicts. As correcting elders transgresses cultural norms in some cultures, young adults carefully consider the severity of misinformation, their expertise on the topic, rapport with the relative, and the relative's potential reactions before approaching them~\cite{Malhotra-2023}. Moreover, they carefully choose the appropriate channel to \textit{politely} correct relatives so that their act of correction does not reflect poorly on their family and does not offend the relatives~\cite{Pearce-2022, Malhotra-2022}.

Research shows that users are more likely to correct misinformation in groups with weak social ties~\cite{Kalogeropoulos-2023}. \citet{Varanasi-2022} show that in neighborhood groups in India, knowledgeable community members and admins often correct COVID-related misinformation \textit{politely} and \textit{indirectly}. However, in work-related groups in India, higher management \textit{directly} calls out the offender for sharing any religious or political propaganda, or irrelevant messages~\cite{varanasi2021tag}. 

Although these studies provide useful insights into social corrections on WhatsApp, they do not inform much about admins' roles, particularly when WhatsApp has increased admins' moderation power~\cite{admindelete-2022}. 
We extend this scholarship by carefully investigating admins' roles and their attitude towards moderation in different public and private groups in India and Bangladesh.  

\subsection{Care and Control in Content Moderation: Parenting Styles as a Lens} \label{subsec:parenting-styles}
Several HCI scholars have examined the roles of moderators through the lenses of care and control. For example, the power held by volunteer moderators has been likened to that of a dictator, governor, police, and judge~\cite{Matias-2019, Wohn-2019, Gerrard-2020, Seering-2022}. Others have positioned moderators as \textit{``care worker, who keep the community safe''}~\cite{Yu-2020, Ruckenstein-2020} or as \textit{``custodian, who help the community grow''}~\cite{Seering-2022}. Given, the intimate nature of private WhatsApp groups and the ``open-forum'' like nature of public groups, we wanted to examine how admins exercise care and control during moderation. 

Several prior studies have used ``parenting'' as a metaphor to capture how care and control manifest during moderation. \citet{Seering-2022} found that volunteer moderators often view online misbehavior as \textit{``childish''} and try to maintain order with a mix of patient instruction and directness. \citet{gillespie-2018} discussed how users often perceive community moderators' guidelines as the rules of a \textit{``stern parent''}. \citet{Register-2023} used Attachment Theory, typically used to describe parent-child dynamics, to characterize how punishments and rewards by Instagram's moderation algorithm affect users. In line with these scholarly work, we use psychologist Diana Baumrind's typology of ``parenting styles''~\cite{Baumrind1971} as a theoretical lens to characterize the diverse expressions of care and control during volunteer moderation on WhatsApp.


\begin{figure}
    \centering
    \includegraphics[width=0.7\textwidth, trim={1cm 0.4cm 0.7cm 0cm},clip]{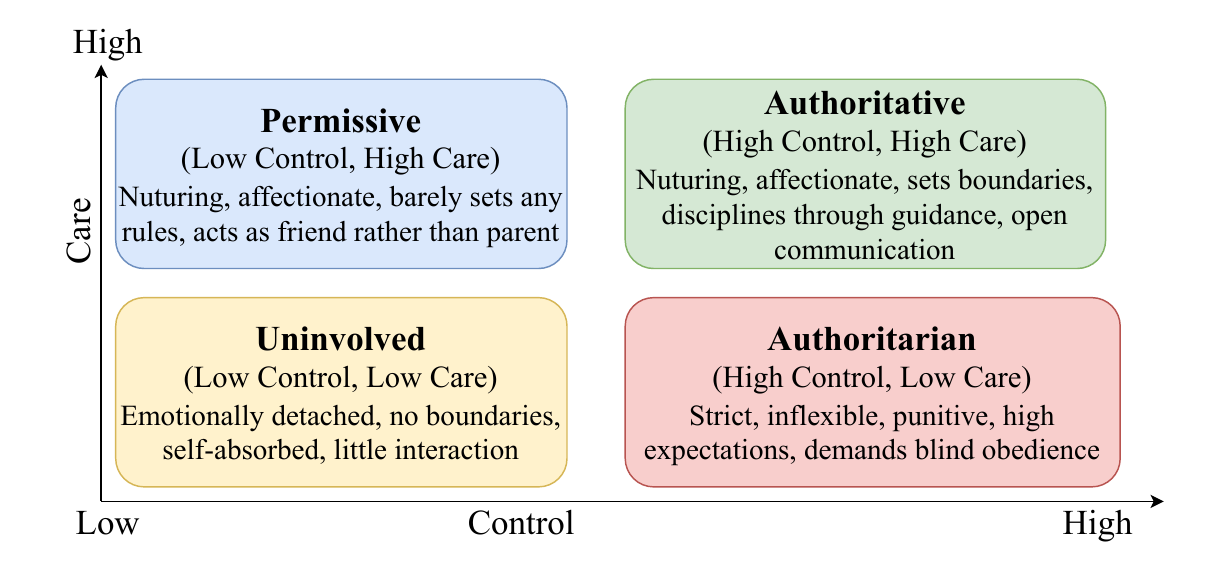}
    \caption{Four parenting styles according to Baumrind's typology.}
    \label{fig:baumrind_typology}
    \Description{A 2D plot of Baumrind's typology. X axis represents Control with values low to high (from left to right). Y axis represents Care ranging from low to high (from bottom to up). Top right segment is Authoritative (high care, high control): Nurturing, affectionate, sets boundaries, disciplines through guidance, open communication. Bottom right segment is Authoritarian (low care, high control): Strict, inflexible, high expectations, punitive, demands blind obedience. Top left represents Permissive (high care, low control): Nurturing, affectionate, barely sets any rules, takes the role of friend rather than parent. Bottom left is Uninvolved (low care, low control): Emotionally detached, self-absorbed, no boundaries, little interaction.}
\end{figure}

Baumrind's pioneering research identified three parenting styles (e.g., authoritative, authoritarian, and permissive parenting) associated with different parenting outcomes~\cite{baumrind1966, baumrind1967, Baumrind1971}. Later \citet{Maccoby_Martin_1983} bridged Baumrind's typology with the two dimensions of parenting: responsiveness (care) and demandingness (control). Here, responsiveness refers to the degree to which parents are sensitive to their children's emotional needs and accepting of their behavior~\cite{Kopko2007, Li2023}. This is described by constructs, such as care, warmth, and support~\cite{power2013}. In contrast, demandingness refers to the degree to which parents control their children's behavior and demand maturity~\cite{Kopko2007, Li2023}. Based on these two dimensions, \citet{Maccoby_Martin_1983} expanded Baumrind's typology and defined four parenting styles: authoritative (i.e., high control and high care); authoritarian (i.e., high control and low care); permissive (i.e., low control and high care); and uninvolved (i.e., low control and low care). 

According to Baumrind's typology~\cite{baumrind1991} (see Figure~\ref{fig:baumrind_typology}), authoritative parents are caring and clearly communicate their rules and disciplinary actions to children. In contrast, authoritarian parents are punitive, less nurturing, and expect children to obey their rules. On the other hand, permissive parents barely impose any rules and act more like friends rather than parents, whereas uninvolved parents remain detached from their children's lives and do not assert any rules. Research shows that authoritative parenting helps children self-regulate and act responsibly~\cite{morris-2007}, whereas children with authoritarian parents often develop high levels of aggression against strict authority~\cite{masud-2019}. Besides, limited moderation by permissive parents often leads to impulsive attitudes and lack of self-regulation among children~\cite{piotrowski-2013}. Similarly, children whose parents remain uninvolved struggle to cope and control their emotions~\cite{kuppens-2019}.

Although Baumrind's typology is based on two-parent, middle-class, and White Western families, it has been applied to different cultural contexts and demographics. Some studies show that the effects of parenting styles are consistent in both individualist and collectivist cultures~\cite{Sorkhabi-2005, pinquart-2018}, including our regions of study: India~\cite{garg-2005, sahithya-2019} and Bangladesh~\cite{karim-2013}. The typology has been useful in modeling complex human relationships around technology use that require both control and care. For example, \citet{valcke-2010} first extended this to children's Internet usage and introduced \textit{``Internet parenting styles."} Since then, the typology has been used to examine parent-child communication online~\cite{zhou2017media}, parental control on children's digital devices and social media usage~\cite{yardi2011social, ching-2017, ren-2022}, and their impact on adolescents' online behavior~\cite{ghosh2018apps, ghosh2018peer, wisniewski2014safety, Jia-2015} along with parents' online sharing practices about their kids~\cite{amon2022sharenting}. Beyond parent-child, the typology has been used to model how the teacher-student relationship influences students' learning outcomes~\cite{yong2019student}. 

\color{changes}We extend the application of Baumrind's typology to examine the roles of WhatsApp admins, given their position demands both care and control to keep the community safe. Prior research has used various metaphors, such as curator, custodian, gardener, teacher, mediator, police, dictator, judge, manager, protector, etc. to capture the diverse roles of volunteer moderators on Facebook, Reddit, Twitch, and Weblog~\cite{Seering-2022, Yu-2020}. For example, \citet{Seering-2022} described nurturing and supportive (caring) moderators as `gardener' and `custodian', whereas the moderators, who prioritize governance and regulation (controlling) were labeled as `dictator' and `judge'. However, being tied to specific roles, these metaphors only capture a singular dimension of care or control. Thus, they may not be able to represent the diverse spectrum of care and control that moderators often exhibit depending on the context. Given Baumrind's typology provides a structured framework to explore the interplay between care and control, we used it to develop a nuanced understanding of how WhatsApp group admins navigate their responsibilities across private and public groups. 

\color{black}
\section{Methods}

To examine how WhatsApp group admins manage their groups and handle problematic content, we performed an in-depth qualitative study with admins based in India and Bangladesh. We define \textit{``problematic content''} as misinformation, hate speech, spam, or anything inconsistent with the group's purpose. We conducted semi-structured interviews with private and public group admins to learn about their experiences of moderation. We also joined public groups to identify instances of problematic content and observe how admins respond in practice. 
Our study protocol received an exemption from the Institutional Review Board of our institution.

\parabold{Participant Recruitment} To recruit WhatsApp group admins, we advertised our study on Facebook, WhatsApp, and X, and requested our contacts to share the study within their social networks. We also took help from our contacts in a grassroots organization based in rural India to recruit admins from rural regions. Through this, we were able to recruit 23 admins from private WhatsApp groups, who 
managed at least one group in India or Bangladesh, and had encountered problematic content in the groups they managed. 
To recruit public group admins, we randomly joined $118$ public groups on various topics (e.g., entertainment, politics, business) in India and Bangladesh that were listed online (e.g., \url{https://whatsgrouplink.com/}). 
We then reached out to the admins of 30 public groups, where we observed group members frequently sharing obvious problematic content (e.g., child porn, monetary scams) during the study period. Through this, we were able to recruit nine public group admins. We coordinated with all our participants via WhatsApp, shared details of our study, and scheduled interviews with them. We continued to recruit admins until the responses reached theoretical saturation~\cite{pandit-1996}.

\parabold{Interviews with Group Admins} We conducted semi-structured interviews with 
32 group admins to learn about their experiences of managing WhatsApp groups. We conducted the interviews in English, Bengali, or Hindi depending on the preferences of the participants. After obtaining informed consent, we asked them in detail about how they became admins in WhatsApp groups, what roles they played, the dynamics among the group members, and their group environment. We then asked them what type of problematic content they experienced in their groups, what were their strategies to handle such content, and the resultant challenges. We also asked them about the support they expected from WhatsApp to better manage their groups. During the interview, some admins voluntarily shared messages from their groups that they found to be problematic. 
After each interview, we revised our questions if needed, stopping when the responses reached saturation. We conducted the interviews either in-person or online. Each interview lasted approximately 40 minutes and was audio-recorded with the consent of the participants. Depending on the geographic location of the participants, we compensated them with 500 INR in India, 500 BDT in Bangladesh, or 10 USD for expats living abroad. 


\parabold{Observations of Public Groups}
Among the public groups we joined, we observed the top $30$ groups in which group members most recently sent messages during the study period. We recorded the group name, group description (if any), number of admins and participants, who could message in the group (either admin-only or all participants), what type of messages were sent in the group, and how admins reacted to problematic content shared in these groups. The analysis was done only for public groups and we did not join private groups due to privacy concerns. 

\parabold{Data Collection and Analysis} We collected around 17 hours of audio-recorded interview data, several pages of notes from our interviews and observations, and 14 forwarded messages that admins perceived to be problematic. 
We translated and transcribed the interviews, notes, and WhatsApp messages into English before iteratively conducting open coding following reflexive thematic analysis as defined by \citet{braun-2006}. Instead of relying on presupposed codes, we let the data guide us. All authors met regularly to discuss the codes that freely emerged from the data, 
did peer debriefing to resolve disagreements while reviewing and updating the codes, and finalized the emerging themes. Our prolonged first-hand engagement with the data helped us establish credibility and reduce coding biases. After multiple passes, we consolidated 59 codes and categorized them under high-level themes, such as the roles of admins, their responses to problematic content, and their expectations from WhatsApp. 

\parabold{Participant Demographics} Table~\ref{tab:demographic} lists the demographics of the 23 private group admins (13 Bangladeshi and 10 Indian) we interviewed. Among them, 19 identified as male, four as female, and one as non-binary. On average, participants were 30 years old (SD: 8 years), aged between 21--52 years. Most of the participants ($n=17$) were from urban areas of Bangladesh and India. We also interviewed three admins from rural India and four Bangladeshi and Indian expats living in North America. All private group admins were highly educated; majority of them ($n=12$) had a master's degree, six had a bachelor's degree, and five were enrolled in a bachelor's program. They pursued various professions, including software development, journalism, entrepreneurship, business, and jobs in government and private sectors, among others. On average, they had been using WhatsApp for almost 7 years (SD: 2.5 years). 

All public group admins ($n=9$) were men and based in India. They were reluctant to share demographic information to protect their identity. Since they were comfortable interacting with us only informally, we did not collect demographic information from them.

\begin{table*}[t]
\caption{Pseudonym, age, gender, profession, and country of the private WhatsApp group admins.}
\label{tab:demographic}
\resizebox{\columnwidth}{!}{
\begin{tabular}{lllll|lllll}
\hline
\multicolumn{1}{c}{Name}                                                                                                                    & \multicolumn{1}{c}{Age}                                                                           & \multicolumn{1}{c}{Gender}                                                                                                  & \multicolumn{1}{c}{Profession}                                                                                                                                                                        & \multicolumn{1}{c|}{Country}                                                                                                                                                            & \multicolumn{1}{c}{Name}                                                                                                             & \multicolumn{1}{c}{Age}                                                                      & \multicolumn{1}{c}{Gender}                                                                                                   & \multicolumn{1}{c}{Profession}                                                                                                                                                                                               & \multicolumn{1}{c}{Country}                                                                                                                  \\ \hline
\begin{tabular}[c]{@{}l@{}}Himadri\\ Ravi\\ Rudra\\ Mehrab\\ Dipali\\ Jawad\\ Labib\\ Munim\\ Nazma\\ Rafid\\ Shaheen\\ Sabbir\end{tabular} & \begin{tabular}[c]{@{}l@{}}26\\ 24\\ 30\\ 40\\ 22\\ 23\\ 52\\ 21\\ 22\\ 22\\ 28\\ 26\end{tabular} & \begin{tabular}[c]{@{}l@{}}Male\\ Male\\ Male\\ Male\\ Female\\ Male\\ Male\\ Male\\ Female\\ Male\\ Male\\ Male\end{tabular} & \begin{tabular}[c]{@{}l@{}}Software Developer\\ PhD Student\\ PhD Student\\ Private Service\\ Student\\ Student\\ Engineer\\ Student\\ Student\\ Student\\ Government Service\\ Engineer\end{tabular} & \begin{tabular}[c]{@{}l@{}}Bangladesh\\ India\\ Bangladesh\\ Bangladesh\\ Bangladesh\\ Bangladesh\\ Bangladesh\\ Bangladesh\\ Bangladesh\\ Bangladesh\\ India\\ Bangladesh\end{tabular} & \begin{tabular}[c]{@{}l@{}}Tawhid\\ Mansi\\ Rizwan\\ Gautam\\ Hamid\\ Arvind\\ Pranjal\\ Priya\\ Kunal\\ Rakesh\\ Tulsi\end{tabular} & \begin{tabular}[c]{@{}l@{}}39\\ 26\\ 42\\ 24\\ 28\\ 27\\ 26\\ 25\\ 40\\ 46\\ 52\end{tabular} & \begin{tabular}[c]{@{}l@{}}Male\\ Female\\ Male\\ Male\\ Male\\ Male\\ Male\\ Non-binary\\ Male\\ Male\\ Female\end{tabular} & \begin{tabular}[c]{@{}l@{}}Entrepreneur\\ Science Journalist\\ Journalist\\ Research Assistant\\ Researcher\\ Political Leader\\ Researcher\\ Private Service\\ News Reporter\\ Business Owner\\ NGO Supervisor\end{tabular} & \begin{tabular}[c]{@{}l@{}}Bangladesh\\ India\\ Bangladesh\\ India\\ India\\ Bangladesh\\ India\\ India\\ India\\ India\\ India\end{tabular} \\ \hline
\end{tabular}
}
\end{table*}

\parabold{Positionality} One of the authors is from Bangladesh and two are from India. The authors who conducted the interviews are native Bengali and Hindi speakers. The authors have been using WhatsApp for more than a decade, are admins of several WhatsApp groups, and have first-hand experience of encountering problematic content in these groups. All authors have extensive experience in conducting fieldwork in India and Bangladesh. Their sociocultural and educational backgrounds are on par with that of the participants. The shared background helped them understand the sociopolitical and cultural dynamics that shape these groups and elicit nuanced responses from the participants. All authors approach HCI from a postcolonial lens~\cite{Irani-2010} and strive to uncover the cultural and sociopolitical epistemologies of the Majority World, whose users dominate online spaces like WhatsApp and shape the nature of governance on such platforms. 

\section{Demographics of WhatsApp Groups}

\begin{table*}[t]
\caption{Demographics of public and private WhatsApp groups in our sample.}
\label{tab:group_details}
\resizebox{\columnwidth}{!}{
\begin{tabular}{|c|c|c|c|}
\hline
Group type & Categories (n)                                                                                                                                                & \begin{tabular}[c]{@{}c@{}}Average number \\ of participants\end{tabular}     & Examples of groups                                                                                                                                                                                                                                                                                                                     \\ \hline
Private    & \begin{tabular}[c]{@{}c@{}}Family (12)\\ Friends (10)\\ Educational (9)\\ Professional (8)\\ Organizational (5)\\ Neighborhood (4)\\ Hobbies (3)\end{tabular} & \begin{tabular}[c]{@{}c@{}}30\\ 20\\ 130\\ 170\\ 110\\ 250\\ 120\end{tabular} & \begin{tabular}[c]{@{}c@{}}Immediate family members and relatives\\ Close friends\\ School, college, or university students\\ Colleagues, employees, engineers, job seekers\\ Religious, political, non-profit, development organizations\\ Local neighbors, communities, villagers\\ Gamers, animal rescue volunteers, entomophiles\end{tabular} \\ \hline
Public     & \begin{tabular}[c]{@{}c@{}}Earning (9)\\ Business (6)\\ Services (6)\\ Educational (6)\\ Entertainment (3)\end{tabular}                                       & \begin{tabular}[c]{@{}c@{}}300\\ 280\\ 320\\ 150\\ 330\end{tabular}           & \begin{tabular}[c]{@{}c@{}}Vacancies, work from home, investments\\ Business opportunity, product sales and promotion\\ Immigration, travel, paid social media promotion\\ Exam preparation, tuition, blogging, photography\\ Sports betting, celebrity fans\end{tabular}                                                                         \\ \hline
\end{tabular}}
\end{table*}

\parabold{Private Groups} The private group admins ($n=23$) in our study managed three groups on average (SD: 3). Most of them ($n=10-12$) were admins of either family or friends groups (see Table~\ref{tab:group_details}). Others managed groups related to their educational institution, 
workspace, neighborhood, or different organizations they were part of. Few participants ($n=3$) also ran personal hobbies and interest-based groups. Private groups varied greatly in group size. Family and friends groups were usually small and close-knit with 20--30 members on average, whereas other types of groups (e.g., professional, work-related) were fairly large with 100--200 participants on average. Private groups were used for various purposes. Family and friends groups were mainly used to stay in touch and share personal updates and photos with close ones. Educational groups were used for study-related discussion and to stay connected with classmates. Organizational and professional groups were used exclusively for work, including performing administrative tasks and monitoring employees' activities. Political leaders, who were admins of their party's dedicated WhatsApp group, orchestrated political campaigns via the group. 

Admins reported that group members actively participated in these groups to share information, opinions, worldviews, and resources. For example, some group members shared information about upcoming events, government notices, and job opportunities. Group members also posted blood donation requests, admission-related questions, and requested advice on technological issues, among others. Regardless of the type, almost all private groups received messages on politics, religion, and issues of national and regional interest. 

\parabold{Public Groups} We examined 30 public WhatsApp groups with various focus areas, ranging from business, earning opportunities, online services, education, and entertainment, among others (see Table~\ref{tab:group_details}). The nine admins we interviewed from these groups managed seven groups each on average. Although most of these public groups were large with 300 participants on average (SD: 205), we barely observed any interaction among the group members. In most groups ($n=25$), group members shared irrelevant and problematic messages that we discuss in detail in Section~\ref{sec:prob_content}. In one-third of the public groups we joined ($n=9$), admins disabled messaging among the group members and only shared messages that were pertinent to the group's purpose, e.g., product details in business groups, visa and immigration services in travel groups, and exam questions in educational groups. 

\section{Findings}
We first outline the paths that led our participants to become admins in WhatsApp groups along with their roles as admins (Section~\ref{sec:admin_role}). We then discuss admins' perceptions of problematic content (Section~\ref{sec:prob_content}) and use Baumrind's typology to categorize admins' responses to such content and the aftermath of their actions (Section~\ref{sec:baumrind}). Finally, we present admins' expectations from WhatsApp to improve handling of problematic content in their groups (Section~\ref{sec:improv}). 

\subsection{\label{sec:admin_role}Who Gets to Become Admins in WhatsApp Groups and What Roles Do They play?}
We found several differences in how admins were selected and what responsibilities they assumed.  

\subsubsection{Becoming an Admin} Majority of the participants ($n=19$) in our sample automatically became admins through WhatsApp's default setting when they created the group. In contrast, only a few participants volunteered to be admin ($n=4$) or were appointed by existing admins ($n=3$). Dipali, an admin of a family WhatsApp group in Bangladesh, shared: 

\begin{quote}
    \textit{``Although my aunt created the group, she became busy with household chores and kids and made me an admin instead. Recently one of my cousins volunteered to be a co-admin to handle the misleading content shared by some relatives.''}
\end{quote}

Different private groups have varying criteria to select admins. In family and friends groups, people became admins due to their existing relationship with the group members ($n=5$). In organizational and professional groups, people with administrative power or high-ranked positions offline (e.g., CEO, managers, supervisors) became admins ($n=6$). In a few educational groups ($n=5$), students who stayed up-to-date with academic events, were socially popular, and demonstrated leadership skills were preferred as admins. Some participants, who  were perceived to be tech-savvy ($n=4$) or active group members ($n=7$), were also made admins. Munim, who was an admin of a neighborhood group in Bangladesh, shared:
\begin{quote}
    \textit{``Earlier only the elders in our housing society could become admins. But, they were not tech-savvy and couldn't understand all the features of WhatsApp. Then they recruited us because I always have Internet connectivity and check the group actively.''}
\end{quote}

Moreover, two groups followed elaborate processes to elect admins through voting or in-person interviews. Even in two friends groups, all group members were made admins so that none could hold power over others. Private groups in our sample often had multiple admins (mean:5, SD: 3.5), while the public groups we observed usually had one admin (SD: 1.5). Admins of private groups often appointed multiple admins so that new group members could be added easily. This was especially helpful in large private groups, where admins were often busy with their own work and preferred to have co-admins to manage the group. Sometimes, the decision to add new admins was strategic. Rakesh, who was an admin of a large religious group in rural India, commented:

\begin{quote}
    \textit{``I made others admin so that they could add their acquaintances to my group instead of forming new groups. This will help my group grow bigger and popular.''}
\end{quote}

However, admins of public groups and few private groups ($n=3$) enjoyed having ownership and control of the group and felt reluctant to add new admins. They feared that the other admin might remove them from the group, seize the power, or add random people to the group. Moreover, in public groups, where members were usually strangers, admins considered it would be \textit{``unwise''} to make unknown people co-admins.

These findings show that apart from the default way of becoming an admin by creating a group, the selection of admins varies depending on the group's needs, dynamics, social hierarchy, and individual's skills.

\subsubsection{The Roles of Admins} 
WhatsApp groups do not have any formal onboarding process and training for admins. Only a few private group admins ($n=4$) reported receiving informal instructions from other admins on how to manage their groups. Thus, in the absence of proper onboarding processes, group management was often left to the discretion of admins. Although several public ($n=6$) and few private group admins ($n=3$) assumed  \textit{``there was nothing to do in WhatsApp groups''} or they were \textit{``too busy to manage the groups''}, most admins actively supervised their groups.

\parabold{Adding Group Members}
Most private group admins ($n=14$) added new members and updated the new phone numbers of existing group members. In contrast, public group admins did not have to bother about adding members as the invite links to join groups were posted publicly. 
Usually private group admins carefully regulated who could join their groups. For example, Dipali declined her aunt's request to add a distant relative to their family WhatsApp group to maintain privacy. Another admin Kunal, who ran a professional group of journalists in rural India, shared his criteria:
\begin{quote}
    \textit{``We don't add any unknown people to our group. We only add those who are affiliated with the press or are able to provide news materials.''}
\end{quote}

A few admins ($n=4$) went through elaborate verification process before adding people to their groups. For instance, in large private groups with weak social ties (e.g., neighborhood, educational, and professional groups), admins often asked for relevant ID and other details to confirm identity. They also asked other group members if they knew the person who wanted to join the group and could vouch for them. Munim further explained the verification process in his neighborhood group:
\begin{quote}
    \textit{``When somebody wants to join the group we ask for their building and apartment numbers. We have a list of contact details for all residents in our housing society. We dial the corresponding apartment to verify if the person actually lives there.''}
\end{quote}


\parabold{Instilling a Sense of Safety and Belongingness}
Several private group admins ($n=8$) felt protective of their groups and tried to create a safe and welcoming space online. For example, Tulsi, who supervised a group of women health workers in rural India, warned women in her group against monetary frauds when such messages were shared in the group. In hobby and interst-based groups, admins often used words of affirmation to engage people in group activities. In addition, admins worked as mediators to resolve online conflicts among group members. Sometimes they handled conflicts offline, particularly when they knew the group members in person. One of the Bangladeshi admins Rizwan shared:
\begin{quote}
    \textit{``A Hindu colleague left our office's WhatsApp group due to hate speech. When we noticed, we decided to apologize to him in person instead of contacting him online given the severity of the matter. After meeting, we requested him to rejoin the group.''}
\end{quote}


\parabold{Addressing Offline Needs}
As private group admins often knew the group members, in several groups ($n=7$) admins assumed offline responsibilities as part of their role. For example, admins in neighborhood and college groups were often tasked with organizing community events as they could leverage their groups for coordination. They also maintained a record (e.g., contact details, professions, blood groups) of all members in case anyone within the group needed help. In job opportunities related group, admins often helped group members by providing referrals upon request even if they did not always know them personally.  


\parabold{Distribution of Workload}
In private groups, admins usually knew fellow co-admins, often met them in person, and were on good terms. Several private group admins ($n=8$) did not bother to differentiate responsibilities among the co-admins because either \textit{``there was not much to do''} or whoever available managed the group. Some private group admins ($n=6$) maintained separate WhatsApp or Messenger chats among themselves to privately discuss the group's logistics. They generally discussed whether something was inappropriate, what actions to take, how to deal with difficult group members, and how to ensure uniformity in their actions. However, the power differential among co-admins dictated these conversations and assumed duties. For example, younger admins in family WhatsApp groups were expected to check with older co-admins before taking any action. Such power hierarchy often led to friction among the admins. For example, Ravi shared an incident from his family WhatsApp group in India:
\begin{quote}
    \textit{``Previous admins had fights because the elderly admin argued that non-blood relative should not send too many messages. But, the younger admin disagreed and was forced to leave the group.''}
\end{quote} 

In a few organizational groups ($n=4$), admins distributed tasks based on their expertise and official position within the organization. For example, as the treasurer of a Bangladeshi graduate student organization, Rudra's task was to post about funding, budget, and membership fees in their organization's WhatsApp group. Tulsi further commented about gendered division of roles: 
\begin{quote}
    \textit{``Since all members in our group are women healthcare workers and the other co-admins are men, it's appropriate that I [a female admin] deal with the group affairs.''}
\end{quote}


\subsection{\label{sec:prob_content}What Types of Problematic Content Affect WhatsApp Groups?}
We now present admins' account of what they perceived as problematic content along with how group members responded to them.

\subsubsection{Admins' Perceptions of Problematic Content} 
Admins had diverse opinions regarding what they thought to be problematic; ranging from obviously harmful messages to often harmless but irrelevant content.

\parabold{Religious Hate Speech and Propaganda}
One-third of the private group admins ($n=8$) reported religious hate speech and propaganda as one of the most common and vicious forms of problematic content they encounter in their groups. Admins in Hindu-dominated groups in India recounted various instances of pro-Hindu and Islamophobic hate speech propagating in their groups 
that they suspected to be originating from right-wing political groups. Gautam, who managed a group of students in his college in India, reported:
\begin{quote}
    \textit{``In our group, people share religiously charged posts that would blame the Muslims for the 2019 Delhi riot, the 2020 COVID pandemic, the 2023 Odisha train collision, or almost anything that might go wrong in this country.''}
\end{quote}
In contrast, Muslim-dominated WhatsApp groups in India received many far-right, pro-Muslim content. Admins mentioned seeing old videos of riots and communal violence (e.g., Hindus burning mosques) being propagated as recent events, conspiracy theories blaming the leftist state government for discriminating against Muslims, or issuing COVID vaccines with pig remnants to hurt the religious sentiment of the Muslims. We noted similar dynamics in WhatsApp groups based in Bangladesh, where Hindus are the religious minority. Arvind commented:
\begin{quote}
    \textit{``Every year during Durga Puja [Bengali Hindu religious festival] there are posts with anti-Hindu sentiment, that would blame the Hindus for disrespecting the Quran, the Prophet, or Muslim women to justify violence against them.''}
\end{quote} 
Similarly, admins reported religious extremism in small-scale pro-Hindu groups in Bangladesh.

\parabold{Political Propaganda}
More than half of the admins in private groups ($n=13$) considered political messages to be problematic as they often led to fights among group members. Admins routinely dealt with pro-political content shared in favor of the ruling party, misinformation and propaganda against the opposition political parties and foreign countries, and conspiracy theories blaming the government for wrongdoings and corruption, among others. Some admins feared that the government might track anti-government messages on WhatsApp and accuse them of spewing distrust. Rizwan shared his thoughts:
\begin{quote}
    \textit{``People think writing on WhatsApp is safe. I doubt if WhatsApp's encryption would work in Bangladesh given the country's strict digital law against anti-government content. The government might trace such messages on WhatsApp and accuse admins.''}
\end{quote}

\parabold{Misinformation, Fake News, and Spam}
Half of the private group admins ($n=12$) reported misinformation and fake news to be pervasive in their groups. They gave examples of COVID-related misinformation that denied the impact of COVID, discouraged people from taking vaccines, and promoted fake cures. Admins also encountered AI-generated fake videos, viral clickbait content from other social media platforms, and old pictures and videos being circulated as current news. They complained about spam messages promising lotteries and shopping discounts as these derailed the conversation in the group and compromised the security of group members. Dipali shared an experience from her college WhatsApp group:
\begin{quote}
    \textit{``After clicking on a spam link shared in the group, many group members' Facebook accounts got hacked. Some girls' sensitive photos were leaked and when we informed our teachers, they filed a cybercrime police complaint. The police interrogated everyone, recovered the hacked accounts, and asked the admins to disable the group.''}
\end{quote}

Although we observed a large presence of monetary scams, fraudulent cryptocurrency schemes, illegal betting services, and sales of illicit arms and hacked social media accounts in public groups, few admins ($n=3$) acknowledged that these messages were problematic.

\parabold{Differential Notions of Problematic Content}
We found that admins treated some types of content differently depending on the context and group dynamics. For instance, admins disliked the use of profanity in educational groups but were fine with group members cursing in friends groups. A few admins ($n=5$) considered travel photos, pleasantries, and birthday messages to be \textit{irrelevant} and \textit{annoying} but felt awkward about forbidding casual conversations. Moreover, admins of political groups prohibited sharing positive news about their opposing parties. Two admins found jokes and sarcasm to be problematic, particularly when shared out of context. Mehrab recounted an incident from his professional group in Bangladesh:
\begin{quote}
    \textit{``Recently while everyone was paying respect to a deceased colleague, someone shared a joke without paying attention to the ongoing conversation. This is \textit{insincere} and displays a lack of common sense.''}
\end{quote}

Several private group admins ($n=6$) disapproved of sharing adult content in family, professional, and other groups that have female group members. In contrast, public groups were filled with child pornography, sex chat and video services, and pornographic images and videos with little-to-no actions from admins.

\subsubsection{User Reactions to Problematic Content} Users in public groups seldom responded to problematic content shared in their groups. Whereas, private group admins reported diverse reactions from users towards problematic content. For example, group members were often divided whether certain types of content, e.g., religious messages or curse words should be allowed. Admins shared that when members felt overwhelmed by the sheer volume of problematic content propagating in their groups, they opted to leave the group rather than calling out the offense. Usually group members ignored political or communal propaganda to avoid conflicts with people in their offline social sphere. Hamid shared his experience in a college WhatsApp group:
\begin{quote}
    \textit{``Communal hate speech has been normalized in India over the years and none has the time or energy to protest such content. Most group members just care about staying connected with college friends instead of constantly arguing with them.''}
\end{quote}

Moreover, in a few family groups ($n=4$), elderly members strongly believed COVID related misinformation and blindly forwarded them. In many private groups ($n=10$), members openly supported communal propaganda and hate speech against minorities. Priya, who ran a private group to share internship and job opportunities in India, reported:
\begin{quote}
    \textit{``During COVID many group members blamed Muslims for the rise of COVID in India. This triggered not only Muslim but other considerate group members from different religions, who decided to give up networking opportunities instead of being in groups that discriminated against people for their religion.''}
\end{quote}

On rare occasions, group members contested problematic content. This occurred particularly in private groups in which group members had relatively weaker social ties. For example, Pranjal shared that a doctor in his workspace group called out a message on Cancer treatment as misinformation and encouraged group members to check sources before sharing health-related content (Figure~\ref{fig:mod-strategy}(A)). In a few educational and professional groups ($n=4$), where admins did not intervene at all, some group members resisted religious hate speech and political propaganda, but this often led to heated arguments among group members. In friends groups, sometimes group members reacted with \textit{haha} or \textit{angry} emojis on spam messages and warned others against clicking on spam links. Nazma, an admin of her university classmates' group in Bangladesh, commented:
\begin{quote}
    \textit{``Sometimes people intentionally share phishing links in the group. If I don't notice, other classmates will call them out as spams and question the group member who shared that content.''}
\end{quote}


\subsection{\label{sec:baumrind}How do Admins Respond to Problematic Content in WhatsApp Groups?}
We now present our analysis of admins' responses to problematic content through Baumrind's typology of parenting styles, with particular focus on how admins exercise \textit{care} and \textit{control} during volunteer moderation. Based on admins' responsiveness to the needs of the group members and demandingness to control their behavior, we describe four moderation styles: authoritative, authoritarian, permissive, and uninvolved. We present these styles as general attitude that we observed broadly among admins in public and private groups in India and Bangladesh, and also in groups with different levels of social ties (strong to weak), instead of a rigid categorization (see Table~\ref{tab:admin_mod}).

\begin{table}
\caption{Characteristics of different parenting styles as per Baumrind's Typology and how they manifest in WhatsApp group admins' attitude towards moderation.}
\label{tab:admin_mod}
\resizebox{\columnwidth}{!}{
\begin{tabular}{|l|l|l|l|l|}
\hline
                                                                                         & \multicolumn{1}{c|}{Authoritative}                                                                                                                                                                                            & \multicolumn{1}{c|}{Authoritarian}                                                                                                                               & \multicolumn{1}{c|}{Permissive}                                                                                                                                & \multicolumn{1}{c|}{Uninvolved}                                                                                                                                                          \\ \hline
\begin{tabular}[c]{@{}l@{}}Key\\ characteristics\\ in Baumrind's\\ typology\end{tabular} & \begin{tabular}[c]{@{}l@{}}Warm, responsive, assertive,\\ high expectations, clear \\ standards, democratic,\\ communicative\end{tabular}                                                                                     & \begin{tabular}[c]{@{}l@{}}Limited warmth,\\ rigid, forceful,\\ adheres to rules, \\ punitive, autocratic\end{tabular}                                           & \begin{tabular}[c]{@{}l@{}}Low expectations,\\ few rules, indulgent,\\ accepting, lenient, \\ avoids confrontation\end{tabular}                                & \begin{tabular}[c]{@{}l@{}}No expectations,\\ absent, passive,\\ neglectful, \\ competing priorities\end{tabular}                                                                        \\ \hline
\begin{tabular}[c]{@{}l@{}}Attitude of\\ WhatsApp\\ group admins\end{tabular}            & \begin{tabular}[c]{@{}l@{}}Cares for group members'\\ wellbeing, sets clear group\\ rules, takes different\\ disciplinary measures to\\ moderate harmful content,\\ gives explanation for\\ moderation decisions\end{tabular} & \begin{tabular}[c]{@{}l@{}}Enables admin\\ only messaging,\\ mutes everyone \\ for the offense of\\ a select few, \\ suppresses group\\ interaction\end{tabular} & \begin{tabular}[c]{@{}l@{}}Reluctant to moderate\\ to avoid conflict and\\ maintain social\\ hierarchy, gives\\ offenders benefit of \\ the doubt\end{tabular} & \begin{tabular}[c]{@{}l@{}}Too busy to manage \\ group, ignores group\\ due to excessive \\ problematic content, \\ lack of moderation \\ draws more \\ problematic content\end{tabular} \\ \hline
\begin{tabular}[c]{@{}l@{}}Manifestation\\ in WhatsApp\\ groups\end{tabular}             & \begin{tabular}[c]{@{}l@{}}Mostly educational, \\ organizational, and \\ professional groups (n=15),\\ few family (n=5) and \\ friends groups (n=2)\end{tabular}                                                              & \begin{tabular}[c]{@{}l@{}}Mostly public\\ groups (n=9),\\ very few family\\ groups (n=2), one\\ professional group\end{tabular}                                 & \begin{tabular}[c]{@{}l@{}}Mostly family and\\ friends groups (n=8),\\ very few professional\\ groups (n=2)\end{tabular}                                       & \begin{tabular}[c]{@{}l@{}}Mostly public\\ groups (n=21),\\ very few private\\ groups (n=3)\end{tabular}                                                                                 \\ \hline
\end{tabular}
}
\end{table}

\subsubsection{Authoritative} \citet{baumrind-1972} described authoritative parents as warm but firm. Authoritative parents encourage their children to be independent, but maintain limits and controls on their actions. They clearly communicate their expectations to children and provide reasoning for their actions~\cite{Kopko2007}. All these lead to positive development outcomes among children~\cite{Li2023}.

We observed this approach mainly in groups ($n=15$) with weaker social ties (e.g., educational, organizational, and professional groups), where people knew each other but maintained a quasi-formal relationship. Admins in some family ($n=5$) and a couple of friends groups ($n=2$) occasionally exhibited this style. As characteristic of authoritative style, these admins took care of members' needs and emotional well-being. Tawhid, who used WhatsApp group to run an online magazine in Bangladesh, shared:

\begin{quote}
   \textit{``Whenever there is communal violence against minorities, I try to deal it with care. I personally message my Hindu employees and ask how they are doing. If they need mental health support or work-related accommodations, I try to provide them.''}
\end{quote}

Authoritative admins supported group members' right to freedom of expression and considered group members \textit{``mature enough''} to behave responsibly. Often, they maintained a balance between being authoritative but yet approachable. 
Mansi shared her view about moderating an animal rescue WhatsApp group in India:
\begin{quote}
    \textit{``We want to bond with the volunteers, who are helping with our cause. However, we need to manage boundaries and establish some sort of command so that we can run the group efficiently.''}
\end{quote}

Depending on the group dynamics, authoritative admins adopted different techniques to control problematic content in their groups, which we outline below.

\parabold{Group Rules} Authoritative admins created group rules to maintain organizational integrity, professional etiquette, and protect group members from harmful content. The rules discouraged the sharing of fake news, adult content, hate speech, spam, gendered harassment, and disrespecting others. Few admins ($n=3$) also disapproved of religiously and politically charged messages. Admins either listed the rules in the group description, or shared them in the group chat or via direct messages when new members joined the group or someone violated the group policy. However, group members did not always follow the rules, which required reactive measures from the admins.

\parabold{Debunking and Fact-Checking} A few admins ($n=4$) used Google, YouTube videos, journalistic and scientific sources 
to fact-check content. However, group dynamics affected how admins informed the group members of the fact-checked content. In groups with peers, admins directly debunked messages in the group chat and asked the offender to review the fact-checking links. However, in family groups, younger admins engaged in a deliberate dialogue to  \textit{respectfully} convince older relatives not to share misinformation by explaining how it would negatively impact group members. Priya described her approach in family group:
\begin{quote}
    \textit{``We don't try to prove elderly relatives wrong when they share health related misinformation. Instead, we give them links and statistics from WHO and suggest to review those to inform themselves better.''}
\end{quote}

In contrast, admins were comfortable correcting younger family members. In professional and organizational groups, admins reprimanded their subordinates for spreading misinformation. However, when the offender held higher position (e.g., a senior colleague), admins messaged them privately so that they do not feel challenged or disrespected in front of other group members. 

Several admins ($n=11$) pointed that when they debunked misinformation and spams, offenders often excused themselves, claiming they had not checked the content before sharing or were unaware of it altogether, speculating that it might have happened when a virus infected their phone or while their children were using it. Sometimes offenders became defensive to deflect the blame. Ravi shared his experience:
\begin{quote}
    \textit{``An elderly relative shared fake ayurvedic [herbal] cures of COVID in the family group. When we tried to debunk, he became defensive, called the scientific articles propaganda, and claimed the ayurvedic remedies had cured someone they knew personally.''}
\end{quote}

\begin{figure*}[t]
    \centering
    \includegraphics[width=\textwidth]{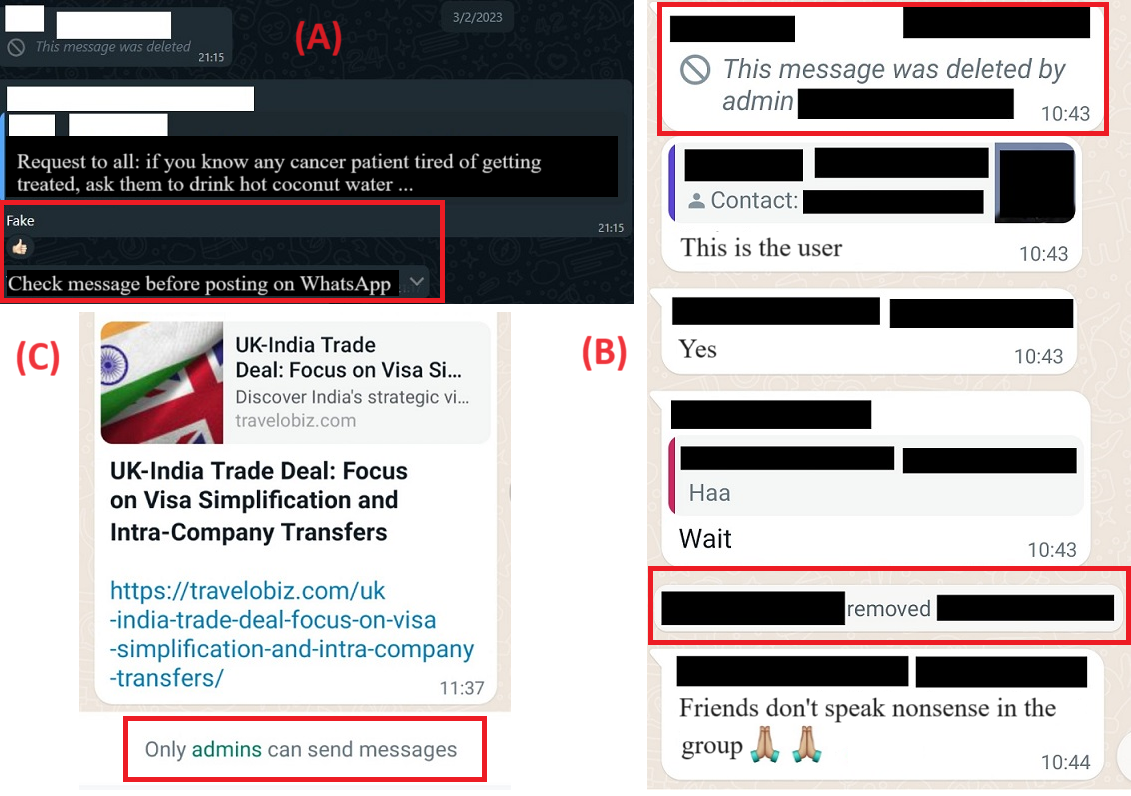}
    \caption{(A) Group member deleted their message on fake cures of Cancer after a group member debunked it. (B) Admin removed both the spam message and the spammer and discouraged group members from sharing irrelevant messages. (C) Admin enabled admin-only messaging in a service related public group. [All messages have been paraphrased in English to protect user privacy.]}
    \label{fig:mod-strategy}
    \Description{(A)Screenshot of a WhatsApp chat: This message was deleted. Admin replied ``Fake'' to a group member's message on fake cures for cancer. Admin asked a group member to check before sending messages to the group. (B) This message was deleted by admin [redacted admin's number]. The admin removed the group member for sending spams. Admin requested to not send irrelevant messages in the group. (C) Admin shared a link on US visa. A label in the bottom said, ``Only admins can send messages.''}
\end{figure*}

\parabold{Content Removal} Most private group admins ($n=18$) adopted a zero-tolerance approach to remove religious hate speech, political propaganda, adult content, and spams in their groups (Figure~\ref{fig:mod-strategy}(B)). When we asked admins about their rationale behind deletion, they expressed that they prioritized the group's welfare over displeasing a single user. They argued that these messages, if stayed longer, could potentially harm group members, hurt their sentiments, and spread more. 

In groups with weaker social ties, authoritative admins requested the offender \textit{privately} either via phone call, WhatsApp, or Messenger to delete the post explaining why it was problematic. If the offender did not respond, then admins tagged them \textit{publicly} in the group chat and asked to remove their post. 
Recently after WhatsApp enabled admins to delete anyone's messages in the group~\cite{admindelete-2022}, several admins ($n=7$) took advantage of this to directly remove harmful content without waiting for users to remove them. Dipali shared an incident from her family WhatsApp group:
\begin{quote}
    \textit{``A female relative had shared an adult content and didn't respond upon contacting. When my younger cousins queried, I asked them to ignore. Ultimately, I deleted the video myself to prevent elderly relatives and younger members from seeing it.''}
\end{quote}

Many admins ($n=16$) received positive responses from the offenders, who readily accepted their mistakes, apologized to the group, deleted their content, and promised to be considerate in the future. However, some admins ($n=6$) reported facing accusations, harassment, and bullying from the offenders for moderating hate speech and political propaganda. Admins were often called \textit{``anti-democratic''} or \textit{``government agent.''} Gautam also faced similar criticism when he condemned Islamophobic hate speech in his college group:
\begin{quote}
    \textit{``I was shocked to see my professors sharing misleading anti-Muslim videos of riots. When I tried to resist, many group members called me names and labeled me as a JNU sympathizer.''} [Jawaharlal Nehru University (JNU) is a public university in India with a strong center-left political foothold.]
\end{quote}

\parabold{Member Removal} Many admins ($n=12$) felt compelled to remove group members, who continued sharing hate speech, spam, and adult content despite multiple warnings. Gautam shared an example from his college group during the 2020 Delhi riot in India:
\begin{quote}
    \textit{``One group member was continuously sharing communally charged gore killing videos of the riot without any verified sources. As these videos were destabilizing the group, we asked him to stop but he didn't listen. Then we permanently removed him.''}
\end{quote}

In some cases, admins gave offenders a second chance and added them back to the group after they apologized and reflected on their behavior. One of the Bangladeshi admins Jawad commented:
\begin{quote}
    \textit{``We blocked a fellow classmate as he was repeatedly spamming the group with links of fake apps. After he clarified that his phone was compromised and apologized, we added him back to the group.''}
\end{quote}

\parabold{Offline Mediation} In a handful of cases ($n=4$), admins dealt with the offenders offline, particularly when their harmful behavior persisted for long. In family groups, admins often requested help from other influential family members, who were in a position to talk to the offender out of sharing misinformation and spams. Ravi shared:
\begin{quote}
    \textit{``One of my uncles was continuously spamming our family WhatsApp group with different health related misinformation. I requested my father to talk to him in person and after that he stopped spamming.''}
\end{quote}

To handle extreme cases, admins often coordinated among themselves to counsel repeat offenders offline, understand their point of view, and explain why their behavior was wrong. Occasionally, admins adopted a format similar to \textit{``public hearing''} to create social pressure on the offender. Rafid, who supervised a college WhatsApp group in Bangladesh, shared:
\begin{quote}
    \textit{``Following several exchanges of communal hate speech in the group, we called group members in the college field including the offenders. There we discussed the issue at length and the offenders pledged they wouldn't repeat this again.''}
\end{quote}

\subsubsection{Authoritarian} \citet{baumrind1991} described authoritarian parents as highly controlling and showing little warmth towards their children. They expect blind obedience from their children and punish harshly, believing that discipline will make the children resilient. Therefore, children often become highly rebellious and develop low self-esteem and behavioral issues~\cite{Li2023}.

Nearly one-third of the public groups we joined ($n=9$) imposed authoritarian moderation. We also noted this in two family groups and a professional group. These admins enforced \textit{admin-only messaging}, i.e., disabled messaging for all members except themselves (Figure~\ref{fig:mod-strategy}(C)). They justified such measures were necessary to manage the inflow of misinformation, spam, and hate speech. Although not all group members were at fault, authoritarian admins chose to mute everyone instead of only disciplining the offenders. Several public group admins we talked to preferred admin-only messaging because it saved their effort to filter harmful content and protected their mental well-being. They prioritized their interest at the expense of group members' freedom of speech. Vivek, an admin of a public gambling group, commented: 
\begin{quote}
    \textit{``I cannot remove the spammers because they are my customers and bring me money. Instead, I can cut spamming easily by enabling admin-only messaging.''}
\end{quote}

The one-way communication, where only admins can send messages, created a power imbalance as group members had no means to hold the admins accountable. For example, in a family group of near and distant relatives, where admin-only messaging was enforced, group members had to send their messages to the admins first, who then curated what messages could be posted in the group. People felt annoyed because they could no longer exchange greetings and share personal updates instantly. Moreover, admins were not always available to check the messages that group members sent to them, and thus time-sensitive content went unnoticed. Priya described similar challenge in a group created to share job opportunities: 
\begin{quote}
    \textit{``Recently admins were on vacation and could not check DMs from group members daily. Someone sent us a good internship opportunity, but by the time we checked, the deadline was over. It was disappointing, since no one in the group could benefit from it.''}
\end{quote}

Although group members requested admins to temporarily lift the restriction on messaging, authoritarian admins ignored such requests. This decreased group members' participation and many left the group since it no longer fulfilled their needs and restricted the exchange of ideas. Some group members even threatened to create separate groups. Yet, admins preferred the status quo saying they were \textit{``too busy to manage the groups full-time.''}

\subsubsection{Permissive} \citet{baumrind1991} defined permissive parents as nurturing and warm. They set very few rules and are lenient if children break them. They do not like to say \textit{`no'} to their children and usually allow them to have their own way~\cite{Kopko2007}. We noted this attitude primarily among admins in family and friends groups ($n=8$), where people shared strong social ties offline. Although an exception, a couple of admins ($n=2$) in professional groups also showed reluctance to moderate content mainly to avoid conflict with group members, who were part of their social circles. 
Thus, permissive admins did not counter religious or political propaganda to avoid offending group members. Ravi emphasized:
\begin{quote}
    \textit{``I do not respond to political forwards in my family WhatsApp group because it's very difficult to change people, where they have already made up their mind.''}
\end{quote}

In family and professional groups, there were unspoken rules that no one would moderate content from people higher up in the social or professional hierarchy (e.g., older relatives and senior colleagues). Both in India and Bangladesh, correcting or contesting elders is considered \textit{disrespectful} and goes against social norms. Although the unwillingness to moderate due to hierarchy between admins and group members might appear contradictory to usual parent-child power dynamics, research shows that aging parents in collectivist families yield to adult children's decisions to maintain family harmony even though children's actions might contradict parents' preferences~\cite{pyke-1999}.

On the other hand, admins in friends groups did not want to assert themselves when their friends shared fake news arguing that \textit{``WhatsApp forwards aren't serious.''} Admins often gave offenders \textit{``benefit of the doubt''} and gave them a pass assuming the spams were caused by \textit{``some issues in the group member's phone.''} Rizwan, who managed a group of university classmates, commented:
\begin{quote}
    \textit{``Many allow their children to play with phones and the kids may mistakenly click on some link. Then, virus automatically spams the contacts and groups through that person's WhatsApp account.''}
\end{quote}

Few permissive admins assumed that group members shared casteist and religious hate speech to vent out their frustrations, and therefore, they ignored those messages to protect free speech. Pranjal commented:
\begin{quote}
    \textit{``Recently in our group, my colleagues were discussing how lower caste students aren't fit for engineering after a student committed suicide. I know some people feel frustrated thinking DEI initiatives that support lower caste students work against merit. How can I stop my colleagues from feeling deprived and penalize them for their thoughts?''}
\end{quote}

As permissive parenting leads to lack of self-control among children~\cite{Li2023}, permissive admins' lack of moderation resulted in \textit{eco-chambers} in close-knit groups in which hateful content flourished. 

\subsubsection{Uninvolved} Baumrind~\cite{baumrind1991} described that uninvolved parents are indifferent to the emotional and social needs of their children and do not often go beyond fulfilling children's basic needs. They barely set any rules, expectations, or boundaries for the children. They may adopt this style out of frustration due to their own problems (stress, abuse, etc.) and may give up parental authority~\cite{Kopko2007}. As a result, children are likely to suffer from impulsive attitude, aggression, and addiction~\cite{Li2023}.

Majority of the public groups ($n=21$) we observed were completely neglected by the admins. Among the nine public group admins we talked to, many ($n=6$) struggled to recognize their groups and had no idea that the previous admin had arbitrarily made them an admin before leaving the group. Moreover, they were ill-prepared to handle the onslaught of mass spamming done by group members. To avoid spam, several admins ($n=5$) either muted or archived their groups and even uninstalled WhatsApp. Thus, being unsupervised, these groups attracted all kinds of problematic content and reinforced a negative feedback loop. Ashok, who created a public group to share job opportunities in India, denied the harms of monetary scams posted in the group and commented:
\begin{quote}
    \textit{``Not everything is fake. People could benefit from these posts by earning money.''}
\end{quote}

When we probed admins why they did not delete the group or make someone else admin if they did not have time to manage the group, some of them said that they might utilize the group later for business purposes or making money. Vivek shared his plan about his public gambling group:
\begin{quote}
    \textit{``Now that the group is large I may sell it. Whoever buys the group can earn money by posting `earn from home' messages. Even if 200 people pay him, he can earn a lot.''}
\end{quote}

In contrast, few private group admins ($n=3$) in our sample exhibited traits of uninvolved moderation. They excused themselves saying that \textit{``they were too busy with work''} or \textit{``it was too difficult to check the large volume of incoming messages.''}

These findings show that public and private groups not only differ in terms of group's purpose, composition, and size but also vary depending on how admins moderate them. However, these moderation styles are not a rigid profiling of WhatsApp group admins because few admins exhibited multiple moderation styles in a single group, based on the offender's identity, the frequency and severity of the offense, and reactions from other group members. For example, Priya initially moderated hate speech in their professional group (authoritative) but eventually had to disable messaging (authoritarian) due to the proliferation of hate speech. Moreover, depending on the group dynamics, same admins would moderate different types of groups differently. For example, Hamid actively resisted communal content in his college group (authoritative) but avoided correcting misinformation in his family group (permissive). Additionally, different admins moderated similar types of groups differently. For example, we observed various degrees of permissive, authoritative, and authoritarian styles in family groups managed by different admins.

\subsection{\label{sec:improv}How Can WhatsApp Support Admins in Volunteer Moderation?}
When we asked private group admins about who should handle moderation in WhatsApp groups, many of them ($n=11$) absolved the platform of the responsibility. They thought that only admins should be held responsible for curtailing problematic content in their groups. Few admins ($n=3$) felt that both WhatsApp and group admins should assume the responsibility for moderation, but they expected admins to handle most of the work. When we probed admins why they did not expect WhatsApp to lead moderation, Hamid commented:

\begin{quote}
    \textit{``Implementing moderation may not satisfy WhatsApp's interest as a platform. Most people get attracted to WhatsApp because there are heated debates in close groups. WhatsApp needs unsupervised controversy to fuel participation in its groups.''}
\end{quote}

Below we discuss admins' expectations from WhatsApp to improve content moderation.

\subsubsection{Mixed Feelings about Automated Moderation}
Private group admins had mixed feelings about automated moderation tools. A couple of them ($n=2$) told us they had no idea whether automated moderation was possible on WhatsApp. Few of them ($n=3$) opposed large-scale automated moderation, fearing lack of cultural sensitivity in moderation outcomes. Rizwan expressed his concerns:
\begin{quote}
    \textit{``In a family group if someone jokingly talks about `beating' and the word is blacklisted, then their message may automatically be removed. People may not understand such auto-removal and it may negatively affect the relationship among group members.''}
\end{quote}

Hamid further pointed out that without addressing users' intent, such technocentrism would provide a \textit{``patching solution''} as people who willingly spread disinformation would move to a different platform or find workarounds. However, some admins ($n=7$), who did not know how end-to-end encryption works, were open to automated platform-mediated moderation thinking it would neither violate encryption nor user privacy if user's personal data is not used. They wanted WhatsApp to automatically detect and remove or shadowban messages that contain spam links and offensive language. Admins wanted to be able to define blacklisted keywords for their groups so that group members receive automatic warnings if their messages contain any of those words. Moreover, they wanted a WhatsApp bot to automatically fact-check and verify posts that were either forwarded many times or reported by multiple users, saying \textit{``it's not feasible to manually check every group message.''}

Admins felt such automated moderation would be appropriate for large public groups, where people had weak social ties and strangers shared problematic content without facing any consequences. Others recommended making auto-moderation optional so that admins could enable or disable it if needed and review the flagged posts to make decisions. 

\subsubsection{Moderation Related Support}
Several admins ($n=7$) complained about how \textit{``WhatsApp admins don't have any power or moderation tools''} compared to those on other platforms. Given the large volume of incoming messages in some groups, admins wanted dedicated moderators like that in Facebook groups. 
Admins suggested that WhatsApp should inform the offenders of anonymous user reports against their content so that they could learn from feedback. Pranjal commented:
\begin{quote}
    \textit{``Truecaller shows how many users have marked a number as spam. If WhatsApp did something similar, it would discourage people from blindly forwarding spams.''}
\end{quote}

In cases when offenders did not change their behavior, admins wanted the option to mute or ban them \textit{temporarily} so that they could not send any messages to the group for a certain period. In addition, admins struggled to find fact-checking content in local languages and on local issues and requested support from WhatsApp. Ravi described:
\begin{quote}
    \textit{``People don't like to read long articles or watch long fact-checking videos on YouTube that force viewers to see ads. If WhatsApp could partner with fact-checking organizations to make short debunking videos in local languages that could be seen within WhatsApp, then low-literate and non-English speaking group members would be benefited.''}
\end{quote}

Admins also recommended to add credibility indicators to fact-checked misinformation, disable forwarding of such messages, or show pop-ups that discourage users from sharing such content. 

\subsubsection{Logistical Support}
Admins demanded logistical support to manage their groups. For example, similar to Facebook groups, they recommended adding screening questions for people interested in joining groups, where people have weaker or no offline ties. Admins felt that such a feature would reduce their workload of verifying people before adding them to the group. At the time of the interview, WhatsApp had a limit of 2048 characters for group description. Admins wanted WhatsApp to raise the character limit so that they could add group rules for potential members to review before joining. Rizwan remarked that if WhatsApp had `pin messages' feature they could display group rules at the top of the chat so that people could easily access the rules\footnote{WhatsApp has recently released this feature~\cite{pin-msg-2023}.}. Moreover, admins requested a badge for themselves, like that of Facebook admins and Reddit moderators, so that in large groups people can recognize the admins when they post instructions. 

Participants, who also administered Facebook groups, expressed the need for a dashboard similar to the one available to Facebook group admins and Reddit moderators. They wanted WhatsApp to show relevant statistics, such as how many users reported a content, which users sent forwards and spams the most. They thought that these data points would help them easily identify perpetrators and find evidence of past violations. Mehrab mentioned:
\begin{quote}
    \textit{``WhatsApp should give a monthly breakdown of group activities. For example, the number and types of messages, shared links, forwards, photos, and spams, monthly activity log, such as how many new people joined the group, and the list of active users.''}
\end{quote}

These findings show that admins need a range of sociotechnical tools embedded within WhatsApp to identify and manage the spread of problematic content in their groups. 

\section{Discussion}
Using Baumrind's typology of parenting styles, our study explores the dynamics and distinct approaches admins use to display care and control in WhatsApp groups in India and Bangladesh. We outline below how different aspects of volunteer moderation on WhatsApp compare to those of admins and moderators on other platforms and from different cultures. We then discuss design implications to support the diverse needs and dynamics of moderation on end-to-end encrypted platforms like WhatsApp. 

\subsection{Volunteer Moderation Across Platforms}
\subsubsection{Selection and Roles of Admins} Unlike other platforms, most participants in our sample became admins on WhatsApp either by default because they created the group or due to their social position, power, and popularity. Some participants became admins either unexpectedly, voluntarily, or through appointment based on their personal ties, digital skills, and standout group activities, which are also common on other platforms~\cite{Seering-2019, Matias-2019, Wohn-2019, Sultana-2022}. 

We found that similar to other platforms~\cite{Seering-2019, Wohn-2019, Seering-Kairam-2022}, WhatsApp does not provide onboarding for admins. As is common among volunteer moderators on Facebook~\cite{Seering-2019, Malinen-2021, Sultana-2022, Kuo-2023}, Reddit~\cite{Seering-2022}, and Twitch~\cite{Wohn-2019, Seering-Kairam-2022}, private group admins in our sample assumed the responsibility of curating group members, answering questions in the group chat, and creating safe space online. However, there were important differences too. For example, we found that private group admins were expected to provide offline support, e.g., organizing community events as the admin of neighborhood group or resolving conflicts offline as they knew group members in person. 

Different platform-related attributes, such as channel ownership on Twitch, moderation capabilities of admins and moderators as defined by Facebook, and tenure or seniority on Reddit, determine the power of volunteer admins and moderators on these platforms~\cite{Seering-2019}. In contrast, social hierarchy led to the division of roles and power among admins in organizational, professional, and family WhatsApp groups in India and Bangladesh. Office superiors and older relatives enjoyed greater authority over younger and junior co-admins due to their seniority and age, as is common in collectivist cultures. 

\subsubsection{Addressing Problematic Content} As observed in previous work~\cite{farooq-2017, Resende-2019, garcia-2021, Varanasi-2022}, the WhatsApp groups in our sample received widespread misinformation, propaganda, spam, and pornographic content. Since 
WhatsApp can barely intervene due to end-to-end encryption, content moderation on WhatsApp greatly relies on the discretion of group admins, including their values and ideologies. Indeed, our analysis of admins' responses to problematic content through Baumrind's typology revealed variances across different groups (see Table~\ref{tab:admin_mod}). Moderation in public groups in our sample was at two extremes, i.e., admins either did not moderate at all (uninvolved) or disabled messaging in the group (authoritarian). Whereas, group ties and offline relations impacted admins' decision of moderation in private groups, which is uncommon on other platforms.

\parabold{Uninvolved} Uninvolved admins' disregard for moderating problematic content could be compared to that of volunteer moderators in online communities that intentionally promote toxicity, e.g., r/Incels~\cite{gillett-2022}, terrorist and right-wing extremist groups on Facebook~\cite{schwarzenegger-2018} and Telegram~\cite{walther2021}. While volunteer moderators and admins in extremist groups on other platforms allow harmful content, they deploy rules, remove users and comments that challenge their extremist ideologies, and direct users to new channels/ groups when existing ones get flagged or banned. This level of control and involvement contrasts with uninvolved admins' disinterest to manage their groups, leading to spaces where unorganized and unmoderated harmful content flourished.

\parabold{Authoritarian} To control the quality of conversation, volunteer moderators on Reddit, Twitch, and Facebook groups often disable commenting on particular posts~\cite{lo-2018, Juneja-2020, kalsnes-2021} and review posts before they go live~\cite{seering-2017, Kuo-2023}. While these measures limit interaction, they still allow communication compared to disabling group chat \textit{indefinitely}, as authoritarian admins in our sample imposed. Such controls can have different outcomes depending on the context. Prior research shows that during Belarusian protests, admins on Telegram temporarily suspended group chat, following the request of group members to prevent external trolling and spamming~\cite{wijermars2022}. However, ~\citet{williams-2022} pointed that how similar measures undertaken by admins in Indian WhatsApp groups silenced minority views that countered the Hindu nationalist hegemony. Moreover, in our sample, we observed how such authoritarian measure caused frustration among group members and led many to leave the groups. 

\parabold{Permissive} Most admins in family and friends groups in our sample cared about avoiding conflict, not appearing \textit{rude} to elderly group members, and barely controlled problematic content. Irrespective of culture, this sentiment is prevalent among WhatsApp users in India~\cite{Malhotra-2023}, Pakistan~\cite{Pasquetto-2022}, Singapore~\cite{Sheryl-2022}, the US~\cite{feng2022}, and the UK~\cite{scott-2023}. Similar to our observations, these cross-cultural studies point that people avoid social correction in family WhatsApp groups because they lack time, downplay the potential harms, feel discomfort to address religious and political issues, give offenders benefit of the doubt, and perceive correction as futile~\cite{Malhotra-2023, feng2022, Sheryl-2022}. \citet{Malhotra-2023} further explains that in Indian families, people avoid correcting elders on WhatsApp because it could quickly spread to other relatives, suggest poor upbringing by one's parents, and damage personal relationships.

\parabold{Authoritative} Most admins in private groups with weaker social ties (e.g., organizational and professional groups) as well as admins in a handful of family and friends groups in our sample cared about mitigating problematic content. The way authoritative admins in India and Bangladesh nurtured the community, defined group rules, and controlled problematic content bear similarities to that of volunteer moderators on Facebook, Reddit, and Twitch~\cite{Seering-2019, Wohn-2019, Seering-2022, Kuo-2023} as well as to that of admins in family, friends, and work-related WhatsApp groups studied in different countries, such as Kenya~\cite{kigatiira2022}, Nigeria~\cite{udem2020}, Brazil~\cite{garcia-2021}, and the UK~\cite{Chadwick-2023}. 

Contrary to volunteer moderators' lack of transparency on Facebook, Reddit, and Twitch~\cite{Seering-2019, Wohn-2019, Juneja-2020, Kuo-2023}, authoritative admins in our sample offered explanations for penalizing problematic content and dealt with the offenders in person due to existing offline ties with group members. Prior study shows that in private WhatsApp groups, where people know each other, users prefer offline corrections to convey polite tone, observe non-verbal cues and reactions of the person being corrected -- none of which are possible online~\cite{Pearce-2022}. 

Authoritative admins' polite and deliberative moderation of individuals with higher social status (e.g., senior colleagues or elderly relatives) mirrors the cautious strategies adopted by WhatsApp users in collectivist cultures~\cite{Pearce-2022, Malhotra-2022, tutiasri2020}. \citet{Malhotra-2022} noted that due to hierarchies in Indian families, WhatsApp users prioritize the \textit{izzat} (face) of themselves, their families, and the person being corrected, expecting the polite, indirect language will make the offender more open to correction. This also applies to hierarchical and paternalistic office cultures in postcolonial societies like India~\cite{kordyban2016} and Bangladesh~\cite{miah2013}. This power dynamics is evident in the way admins called out and reprimanded their juniors and subordinates publicly for sharing unwanted messages in office WhatsApp groups, which is also documented in prior studies based in India~\cite{Varanasi-2022, varanasi2021tag}. 



\subsection{Improving Volunteer Moderation in WhatsApp Groups} 
Our findings highlight the strengths and weaknesses of different moderation styles practiced by public and private group admins in our sample. While authoritative admins controlled problematic content, they experienced bullying and harassment from the offenders, which soured their relationships with people they knew offline. While authoritarian admins were able to prevent problematic content in their groups, their strict moderation policies (e.g., admin-only messaging) stifled all forms of communication within the group and stripped group members of their agency and freedom of expression. In contrast, lack of moderation by permissive and uninvolved admins led to echo chambers in private groups and turned public groups into a cesspool of problematic content. Given the diverse outcomes of different moderation styles, there is no single moderation tool that might work well for the unique circumstances of different WhatsApp groups. We outline below how to address the diverse needs and moderation styles of different WhatsApp group admins.

\subsubsection{Supporting Authoritative and Permissive Admins} Given people know each other in private WhatsApp groups, it is important to design tools considering the relationship dynamics and how users value those relations.

\parabold{Support to Inform Offenders of Moderation} Many authoritative admins in our study reported being harassed by the offenders following moderation. Hence, admins preferred indirect, respectful language while correcting group members, who they knew in person. This is a common practice among WhatsApp users in collectivist cultures~\cite{Malhotra-2022, Pearce-2022, tutiasri2020}. In contrast, American users prefer objections that make conscientious moral appeal against offenses~\cite{zhao-2023}. However, users in private WhatsApp groups often find it difficult to craft message to challenge misinformation~\cite{scott-2023}. Therefore, WhatsApp can provide templates incorporating localized relational politeness norms for debunking, warning, and explaining moderation decisions to offenders. Given permissive admins hesitate to moderate people who outrank them in social hierarchy, these correction messages will reduce their burden of deciding how to approach the offenders. Moreover, careful choice of words may help de-escalate tension between authoritative admins and offenders, even when the offender is a peer.

\parabold{Adding Frictions and Nudges} 
Prior research shows that social nudges from `ingroup', i.e., those who are from the same race or religion as the offender can reduce hate speech~\cite{munger-2017, siegel-2020}. As observed in family WhatsApp groups in the UK~\cite{scott-2023}, admins in our sample from India and Bangladesh often sought help from elderly relatives to approach offenders of the same age group. However, irrespective of culture, correction in close-knit WhatsApp groups is considered stressful~\cite{scott-2023, Malhotra-2023} and people avoid contesting communal hate speech, religious and political propaganda. Therefore, apart from debunking, WhatsApp could partner with local civil organizations and experts to create social nudges in different formats and languages that would encourage to not spread hatred and division against people from different ethnicity, caste, religion, gender, and political leanings. Since admins in private groups are well-aware of the group dynamics, they could disseminate social nudges by involving individuals, who either share similar background as the offender or have more power over the offender. 

Since WhatsApp users participate in different public and private groups, exposure to problematic content in one group may lead them to share it in other groups and resist admins' moderation decisions. Prior research shows that `nudging' is useful in reducing users' intent to share misinformation~\cite{pennycook-2021}. WhatsApp could nudge users when they try to forward any content and ask them the reason behind sharing. Additionally, WhatsApp's help page provides tips for identifying problematic content~\cite{spam-features}, which they could show as a pop-up when users try to share something viral. These added frictions might dissuade users from impulsive sharing of harmful content.

\subsubsection{Positively Engaging Authoritarian and Uninvolved Admins} As public group admins in our sample were either too strict or lax about moderation, we recommend tools to enable balanced moderation practices in public groups.

\parabold{Authoritative Moderation Tools for Authoritarian Admins}
WhatsApp should provide proper monitoring and preventive tools to help public group admins moderate without resorting to authoritarian measures, which are unfair to innocent members and undermine the group's purpose. Although invite links to join public groups are available online, WhatsApp could introduce screening questions like that of Facebook groups to keep out spamming WhatsApp bots. WhatsApp could use unencrypted account metadata to evaluate user profiles and use automated tools to filter out accounts that give low-quality responses to screening questions. Newly joined accounts could also be restricted from posting or have their initial messages be pre-approved by admins before being posted in the group. 
This might help public group admins immediately identify spammers and remove them from the group. Moreover, WhatsApp could use unencrypted activity metadata, e.g., number of messages and forwards sent by users to spot anomalies and alert admins so that they can review a sample of messages sent by highly active users. Similar to Facebook, WhatsApp can introduce the feature to temporarily suspend repeat offenders~\cite{suspend}, so that they cannot send messages to the group for a certain period. Another option is to `shadowban' the offenders so that no group members can see their messages except admins. These measures might protect innocent users from harmful content without taking away their right to participate in the group and provide recourses to admins to exercise control without being authoritarian.

\parabold{Consequences for Uninvolved Admins}
WhatsApp should intervene when admins in large public groups willingly ignore problematic content, as currently they do not face any consequences. Facebook reduces privileges and reach for the groups that have multiple violations and, in extreme cases, removes the group completely~\cite{alison-2021}. While WhatsApp claims to restrict chat activity in groups where admins violate WhatsApp's Terms of Service~\cite{chat-prvenet-2021}, it's unclear how they do so and we did not come across any groups, where WhatsApp placed such restrictions. WhatsApp could use unencrypted account metadata to detect admins who have stopped using WhatsApp or are no longer active in the group. Then, WhatsApp can send reminders to admins via phone messages or in-app notification to check their groups. If they continue to remain inactive, WhatsApp can disable or remove the group.

\section{Limitations and Conclusion}

We use Baumrind's typology of ``parenting styles" to describe how admins in India and Bangladesh enact care and control during volunteer moderation on WhatsApp. While our findings reveal important distinctions between the composition, dynamics, and workings of admins in public and private groups, we also highlight how moderation in WhatsApp groups qualitatively differ from that of Facebook groups, subreddits, and Twitch channels. Our work has a few limitations. Participants in our study were skewed towards highly educated, urban (in private groups), and male admins (in public groups). Due to small sample size, we might not have discovered all the different ways admins respond to problematic content. For example, volunteer moderators in extremist groups on other platforms allow problematic content but strictly moderate those, who oppose their extremist views~\cite{schwarzenegger-2018, walther2021}. Since we did not have any organized extremist groups in our sample, we could not discover how admins would moderate these groups. 

Given the self-reported nature of qualitative interviews, we could not verify if admins in our study shared problematic content themselves, perceived hate speech or propaganda as harmless due to their own biases, treated the offenders equally, or used their power to suppress minority or opposing views. Since we did not interview group members, we could not assess if the community found volunteer moderation to be useful. Similarly, we could not assess whether users, who left public groups after admins disabled group chat, were indeed innocent group members or spammers. All these could influence the characterization and framing of care and control during volunteer moderation. Furthermore, the moderation styles observed in public and private groups in India and Bangladesh might not apply to groups in other cultures and geographies. More work is needed to unpack the complex dynamics among admins and group members across different demographics. 

\begin{acks}
    We thank Shoaib Dipu and Bharat Nayak for their help in recruiting WhatsApp group admins from Bangladesh and India respectively.
\end{acks}


\bibliographystyle{ACM-Reference-Format}
\bibliography{00_references}

\end{document}